\documentclass[aps,prl,twocolumn,showpacs,byrevtex]{revtex4}
\usepackage{times}
\usepackage{amsmath}
\usepackage{graphicx}
\usepackage{epstopdf}
\usepackage{color}

\newcommand{\kk}{K^+ K^-}

\newcommand{\EE}{e^+e^-}

\newcommand{\ds}{D_{s}}
\newcommand{\dsp}{D_{s}^+}
\newcommand{\dsm}{D_{s}^-}
\newcommand{\dss}{D_{s}^{*}}
\newcommand{\dssp}{D_{s}^{*+}}
\newcommand{\dssm}{D_{s}^{*-}}

\newcommand{\dsss}{D^*_{s0}(2317)}
\newcommand{\dssspm}{D^*_{s0}(2317)^{\pm}}

\newcommand{\dsssm}{D^*_{s0}(2317)^-}

\newcommand{\pidsm}{\pi^{0} D_{s}^-}
\newcommand{\pidspm}{\pi^{0} D_{s}^{\pm}}

\newcommand{\kkpip}{K^+ K^- \pi^{+}}

\newcommand{\pipi}{\pi^+ \pi^-}

\newcommand{\incfig}[2]{\includegraphics[width=#1\textwidth]{#2}}
\def\Journal#1#2#3#4{{#1} {\bf #2}, #3 (#4)}

\def\PLB{Phys. Lett. B}
\def\PRL{Phys. Rev. Lett.}
\def\PRD{Phys. Rev. D}

\def\CPC{Chin. Phys. C}

\def\EPJA{Eur. Phys. J. A}
\def\HEPNP{HEP \& NP}

\def\JHEP{J. High Energ. Phys.}

\def\PTPS{Prog. Theor. Phys. Suppl.}
\begin{document}

\graphicspath{{figure/}}
\DeclareGraphicsExtensions{.eps,.png,.ps}

\title{\boldmath
Measurement of the absolute branching fraction of $\dssspm\to\pidspm$}
\author{
\small
M.~Ablikim$^{1}$, M.~N.~Achasov$^{9,e}$, S. ~Ahmed$^{14}$, M.~Albrecht$^{4}$, A.~Amoroso$^{50A,50C}$, F.~F.~An$^{1}$, Q.~An$^{47,a}$, J.~Z.~Bai$^{1}$, O.~Bakina$^{24}$, R.~Baldini Ferroli$^{20A}$
% ...
\\
\vspace{0.2cm}
(BESIII Collaboration)\\
\vspace{0.2cm} {\it
$^{1}$ Institute of High Energy Physics, Beijing 100049, People's Republic of China\\
% ...
$^{4}$ Bochum Ruhr-University, D-44780 Bochum, Germany\\
% ...
$^{9}$ G.I. Budker Institute of Nuclear Physics SB RAS (BINP), Novosibirsk 630090, Russia\\
% ...
$^{14}$ Helmholtz Institute Mainz, Johann-Joachim-Becher-Weg 45, D-55099 Mainz, Germany\\
% ...
$^{20}$ (A)INFN Laboratori Nazionali di Frascati, I-00044, Frascati, Italy; (B)INFN and University of Perugia, I-06100, Perugia, Italy\\
% ...
$^{24}$ Joint Institute for Nuclear Research, 141980 Dubna, Moscow region, Russia\\
% ...
$^{47}$ University of Science and Technology of China, Hefei 230026, People's Republic of China\\
% ...
$^{50}$ (A)University of Turin, I-10125, Turin, Italy; (B)University of Eastern Piedmont, I-15121, Alessandria, Italy; (C)INFN, I-10125, Turin, Italy\\
% ...
$^{a}$ Also at State Key Laboratory of Particle Detection and Electronics, Beijing 100049, Hefei 230026, People's Republic of China\\
% ...
$^{e}$ Also at the Novosibirsk State University, Novosibirsk, 630090, Russia\\
% ...
}
}

\affiliation{}

\vspace{0.2cm}
\date{\today}

\begin{abstract}
The process $\EE\to \dssp\dsssm+c.c.$ is observed for the first
time with the data sample of 567~pb$^{-1}$ collected with the BESIII detector operating at the
BEPCII collider at a center-of-mass energy $\sqrt{s} = 4.6$~GeV.
The statistical significance of the $\dssspm$ signal is $5.8\sigma$
and the mass is measured to be ($2318.3\pm
1.2\pm 1.2$)~MeV/$c^{2}$. The absolute branching fraction
$\mathcal{B}(\dssspm\to \pidspm)$ is measured as $1.00^{+0.00}_{-0.14}\pm
0.14$ for the first time. The uncertainties are statistical and systematic,
respectively.
\end{abstract}

\maketitle

 \lefthyphenmin=2
 \righthyphenmin=2
  \uchyph=0
The $\dsssm$ meson was first observed at the \textit{BABAR}
experiment via its decay to $\pidsm$~\cite{babar_1,babar_2}; it was
subsequently confirmed at the CLEO~\cite{cleo} and Belle~\cite{belle}
experiments. The $\dsssm$ meson
is suggested to be the $P$-wave $\bar{c}{s}$ state with spin-parity $J^{P} = 0^{+}$. However, the
measured mass $(2317.7\pm 0.6)$~MeV/$c^{2}$~\cite{pdg} is at
least 150~MeV/$c^{2}$ lower than the calculations of a potential model~\cite{model}
and lattice QCD~\cite{lqcd} for such a state. As the
$\dsssm$ is 45~MeV/$c^{2}$ below the $DK$ threshold, it has been
proposed as a good candidate for a $DK$ molecule~\cite{molecule}, a
$\bar{c}{s}q\bar{q}$ tetraquark state~\cite{tetraquark}, or a
mixture of a $\bar{c}{s}$ meson and a $\bar{c}{s}q\bar{q}$ tetraquark~\cite{mixture}.

The $\dsssm$ is extremely narrow, and the upper limit on its width is
3.8~MeV at the 95\% confidence level (C.L.)~\cite{babar_3}.
The only known decay is the isospin-violating mode $\pidsm$,
and no branching fraction or partial width of this mode
has been measured. Theoretical calculations give different values for
the partial decay width $\Gamma(\dsssm\to \pidsm)$ based on different
assumptions~\cite{br_model,br_molecule,br_tetra,br_mix}.
The partial width $\Gamma(\dsssm\to \pidsm)$
is around 30 keV or even as low as a few keV if the $\dsssm$ is a pure $\bar{c}{s}$
state, while it can be enhanced by a hundred keV or even larger in the
molecule picture due to the contribution of meson loops.
Therefore, the partial decay width or the branching fraction is a key
quantity to identify the nature of the $\dsssm$.

In this Letter, we present first observation of $\EE\to
\dssp\dsssm+c.c.$ and the first measurement of the absolute
branching fraction of $\dsssm\to \pidsm$. Throughout the text, the inclusion
of the charge conjugate mode is implied unless otherwise stated.
The data sample, which corresponds to an integrated luminosity of 567~pb$^{-1}$~\cite{luminosity},
is collected
at a center-of-mass (c.m.) energy of 4.6~GeV~\cite{cmsenergy} with
the BESIII detector~\cite{bes3detector} operating at the BEPCII
collider~\cite{bepc2}. In this analysis, a $\dssp$ is reconstructed
via its $\gamma\dsp$ decay with $\dsp$ decaying to $\kkpip$,
 and its recoil mass spectrum is
examined to search for a $\dsssm$ signal. The $\dssp$ tagged sample is
further divided into two subcategories, one with a tagged $\pi^0$ and
the other with no tagged $\pi^0$. By using the numbers of signal
events in these two categories, the absolute branching fraction of
$\dsssm\to \pidsm$ is determined.


In order to determine the detection efficiency and to optimize the
selection criteria, the {\sc geant4}-based~\cite{geant4} Monte Carlo (MC)
simulation software {\sc boost}~\cite{boost}, which includes the geometric description of
the detector and detector responses, is used to simulate $\EE\to
\dssp\dsssm$ at
$\sqrt{s} = 4.6$~GeV with $\dssp\to \gamma\dsp$ and $\dsp\to
\kk\pi^{+}$, and $\dsssm\to \pidsm$ or $\gamma \dssm$.  The $\dsm$ and
$\dssm$ are set to decay inclusively.
The $J^{P}$ of $\dsssm$ is $0^{+}$, so it is
in relative $S$-wave to the $\dssp$, and they are generated uniformly
in phase space. The initial state radiation (ISR) is simulated
with {\sc kkmc}~\cite{kkmc} using a calculation with a precision
better than 0.2\%. The final state radiation (FSR) effects associated
with charged particles is handled with {\sc photos}~\cite{photos}.
To study the possible backgrounds, an inclusive MC sample with an
integrated luminosity equivalent to data is generated. All the
known charmonium transitions, hadronic decays and open charm channels
are modeled with {\sc evtgen}~\cite{evtgen_1,evtgen_2} incorporating
the branching fractions taken from the Particle Data Group~\cite{pdg},
while the QED processes and the unknown charmonium decays are generated with {\sc babayaga}~\cite{babayaga} and
{\sc lundcharm}~\cite{lundcharm}, respectively.

To reconstruct $\dssp$, the $\gamma\dsp$ channel is used with $\dsp$
decaying to $\kkpip$.
 Events with at least three charged track candidates and at
least one photon candidate are selected. For each charged track candidate,
the polar angle $\theta$ in the multilayer drift chamber (MDC) must satisfy
$|\cos\theta |<0.93$, and the distance of the closest approach to the
$\EE$ interaction point is required to be less than 10~cm along the beam
direction and less than 1~cm in the plane perpendicular to the
beam. Particle identification (PID), which uses
both the information from time of flight (TOF) and the specific energy loss
($dE/dx$), is performed to separate kaons and pions. The photon
candidates are selected from showers in the electromagnetic
calorimeter (EMC) with deposited energy greater than 25~MeV in the
barrel ($|\cos(\theta) |<0.8$), or greater than 50~MeV in the end-cap regions
($0.86<|\cos(\theta) |<0.92$). To eliminate showers produced by
charged tracks, the photon candidate must be separated by at least
 20 degrees from any charged track. The time for the shower measured by the EMC from the start of
this event is restricted to be less than 700~ns to suppress
electronic noise and energy depositions unrelated to the event.

All combinations are required to have the invariant masses
of $\kkpip$ and $\gamma\kkpip$ within
$\Delta M_{\kkpip} \equiv \lvert M(\kkpip)- m_{\dsp} \rvert <  16$~MeV/$c^{2}$ and
$\Delta M_{\gamma\kkpip} \equiv \lvert M(\gamma\kkpip) - m_{\dssp} \rvert <  11$~MeV/$c^{2}$, where $M((\gamma)\kkpip)$ is
the invariant mass of the $(\gamma)\kkpip$ system and $m_{\dsp/\dssp}$ is the nominal
mass of $\dsp/\dssp$~\cite{pdg}.
A two-constraint (2C) kinematic fit is performed on the surviving events
with the mass constraints of $\ds$ and $\dss$
to obtain a better recoil mass resolution and to suppress
backgrounds.
 The $\chi^{2}_{\rm 2C}$ from the kinematic fit is required to
be less than 14. All successful combinations in each event
 are kept for further study.

\begin{figure}
\incfig{0.5}{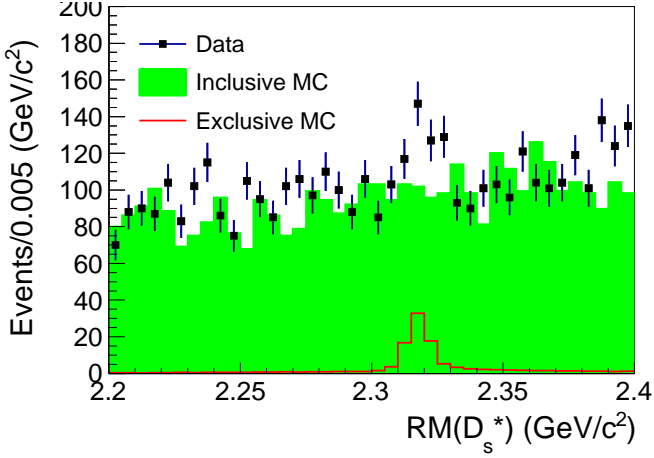}
\caption{(Color online) Distribution of the $\dssp$ recoil mass of the events from data (black dots)
and inclusive MC sample (green histogram), which is normalized according to the integrated luminosity.
 The red curve shows the same distribution
 for $\dssp\dsssm$ events from MC simulation.} \label{fig:mdss_reco}
\end{figure}

After the previously described selection criteria, the recoil mass
distribution of $\dssp$ is shown in Fig.~\ref{fig:mdss_reco}, where
a $\dsssm$ signal can be observed.  The events in the sidebands of
$\dsp$ and $\dssp$ in the sample before the kinematic fit are checked and no signal of $\dsssm$ is observed.
The inclusive MC sample, which does not include production of the $\dsssm$,
matches well with the background from data.
In the inclusive MC sample, the remaining events are non-$\dssp$ events
around the $\dsssm$ peak, including non-$\dsp$ events and
mis-combined $\gamma\dsp$ events, where the $\gamma$ or $\dsp$ could come
from other decay modes of $\dssp$. For the event with a real $\dssp$, such as
$\EE\to\dssp\dssm$ or $\dssp\dsm$, the recoil mass of $\dssp$ is far away from
the $\dsssm$ peak and has no influence in this analysis.
In general, none of the known backgrounds
can form a peak in the signal region. On the other hand,
the technique to measure the absolute branching fraction
$\mathcal{B}(\dsssm\to \pidsm)$ avoids the influence of the unknown
three-body processes $\gamma\dsp\dsssm$ and $\pi^{0} \dsp\dsssm$
even if they exist since they have an identical $\dsssm$ compared to
the signal process $\dssp\dsssm$.

The process $\EE\to \dssp\dsssm\to\dssp \pi^0\dsm$ is
studied via a further $\pi^{0}$ reconstruction with two photons from the remaining showers in the EMC
and $\dsm$ as missing particle.
 If there are more than
two photons, all combinations of $\gamma\gamma\dssp$ are
subjected to a 4C kinematic fit with mass constraints on the $\dsp$,
$\dssp$, $\pi^{0}$ candidates and
a missing $\dsm$, requiring the $\chi^{2}_{\rm 4C}$ to be less than 36.

The requirements on $\Delta M_{\kkpip}$, $\Delta M_{\gamma\kkpip}$, $\chi^{2}_{\rm 2C}$ and $\chi^{2}_{\rm 4C}$
are optimized with MC samples to obtain the best statistical precision of $\mathcal{B}(\dsssm\to \pi^0\dsm)$.
 The $\dssp\dsssm$ signal is generated by
assuming $\mathcal{B}(\dsssm\to \pi^0\dsm) = 0.9$ and
$\mathcal{B}(\dsssm\to \gamma \dssm) = 0.1$ and normalized according
to the number of signal events from data. The background is
taken from a toy MC sample generated by fitting the recoil mass
distribution of $\dssp$ from data. The MC samples are analyzed with the same
procedure as for data to obtain the branching fraction $\mathcal{B}(\dsssm\to \pi^{0} \dsm)$.
The requirements yielding
the smallest relative statistical uncertainty are used in this analysis.

\begin{figure}
\incfig{0.5}{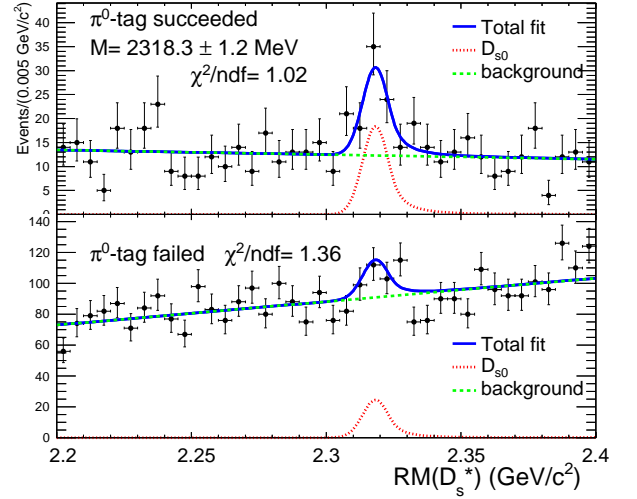}\\
\caption{(Color online) Fit result for data at 4.6~GeV for the two subsamples,
``$\pi^{0}$-tag succeeded" (top) and ``$\pi^{0}$-tag failed" (bottom).
The red dotted and green dashed curves show the fit results for signal and background, respectively,
 while the blue curve shows their sum.} \label{fig:fit_data}
\end{figure}

The $\EE\to \dssp\dsssm$ events are divided in two subcategories:
``$\pi^{0}$-tag succeeded" if at least one $\pi^{0}$ is
tagged and the event passed the 4C kinematic fit, and ``$\pi^{0}$-tag failed" for
the other events.
 The recoil mass distributions of the
$\dssp$ from the 2C kinematic fit of these two subcategories are
shown in Fig.~\ref{fig:fit_data}. These
distributions are fitted simultaneously to measure the branching fraction of
$\dsssm\to \pi^0\dsm$.

The real $\dsssm\to\pidsm$ signal events could be categorized into both subsamples
since the detection efficiency for $\pi^{0}$ is 43.4\%. On the other
hand, potential background
events, such as $\dsssm\to\gamma\dssm$ or other decay channels, could be reconstructed in the
``$\pi^0$-tag succeeded'' sample too.
Therefore, the number of $\dsssm$ signal events in the ``$\pi^{0}$-tag
succeeded" subsample, $N_0$, is expressed as
\begin{equation}
\label{equ:n0}
 N_0 =
  N_{\rm tot}/\epsilon_{\rm tot}\cdot\mathcal{B}\cdot\epsilon_{\rm sig}
 +N_{\rm tot}/\epsilon_{\rm tot}\cdot(1-\mathcal{B})\cdot\epsilon_{\rm bkg},
\end{equation}
where the first and the second terms represent the contributions
from $\dsssm\to \pidsm$ (with a branching fraction of
$\mathcal{B}$) and from the other $\dsssm$ decay mode (with a
branching fraction of $1-\mathcal{B}$), respectively. Here the
other decay mode means the potential peaking background mode
$\dsssm\to\gamma \dssm$, which is expected to be the
dominant mode besides $\pidsm$, and any other decay modes are
considered in the systematic uncertainty.
The $N_{\rm tot}$ is the number of
$\dsssm$ signal events in the full sample (the sum of ``$\pi^0$-tag
succeeded'' and ``$\pi^0$-tag failed'' events), $\epsilon_{\rm tot}$ is
the corresponding detection efficiency for the reconstructed $\dssp$, $N_{\rm
tot}/\epsilon_{\rm tot}$ is the number of produced $\dssp\dsssm$
events, $\epsilon_{\rm sig}$ is the detection efficiency for
$\dsssm\to \pidsm$ events being reconstructed in the
``$\pi^0$-tag succeeded'' sample including the branching fraction of $\pi^0\to\gamma\gamma$~\cite{pdg}, and $\epsilon_{\rm bkg}$ is the
efficiency for non-($\dsssm\to \pidsm$) events to be reconstructed in
the ``$\pi^0$-tag succeeded'' sample.
The efficiencies $\epsilon_{\rm tot}$,
$\epsilon_{\rm sig}$ and $\epsilon_{\rm bkg}$ are obtained from MC
simulations, and are 40.0\%, 17.2\%,
and 5.8\%, respectively.

From Eq.~(\ref{equ:n0}), we derive the absolute branching fraction
$\mathcal{B}(\dsssm\to \pidsm)$ as
\begin{equation}
\label{equ:br}
 \mathcal{B} =
 \frac{N_0-N_{\rm tot}/\epsilon_{\rm tot}\cdot\epsilon_{\rm bkg}}
      {N_{\rm tot}/\epsilon_{\rm tot}\cdot
      (\epsilon_{\rm sig}-\epsilon_{\rm bkg})},
\end{equation}
where the branching fraction $\mathcal{B}$
and $N_{\rm tot}$ are the free parameters in a simultaneous fit to the
recoil mass distributions of the $\dssp$ in
Fig.~\ref{fig:fit_data}, and $N_0$ is calculated using Eq.~(\ref{equ:n0}).

The shape for the $\dsssm$ signal is described with a Crystal Ball
function~\cite{CB} convolved
with a Gaussian function, while the background is parameterized with a linear function.
 The parameters of the Crystal Ball function except for the
mass are fixed to the values from a fit to the MC simulated
$\dssp\dsssm$ sample, in which the $\dsssm$ is simulated with zero
width. The Gaussian function is used to describe the data-MC
difference in mass resolution,
and the standard deviation is taken from
a control sample of $\EE\to D_s^{*+} D_s^{*-}$ at 4.6 GeV.
By reconstructing the $\dssp$ from the process $\EE\to D_s^{*+} D_s^{*-}$,
it is found that the recoiling $\dssp$ signal shape in MC
simulation needs to be smeared by a Gaussian with the standard
deviation of 0.9~MeV/$c^2$ in order to match the data.
The standard deviation of the Gaussian function in the fit to the $\dsssm$ signal
is fixed to this value.

From the simultaneous fit, the total number of $\dsssm$ signal events
is
$115\pm21$, and the number of $\dsssm$ events in the ``$\pi^0$
tag-succeed'' subsample is $46.8\pm9.4$.  The latter event yield is
found to be $49.3$ with a constraint that the branching fraction is no
larger than one.
Using Eq.~(\ref{equ:br}), the
absolute branching fraction of $\dsssm\to \pidsm$ is measured to be
$1.00^{+0.00}_{-0.14}$, with a constraint that the branching fraction cannot
be larger than one. The
statistical uncertainty, 0.14, is estimated by covering 68.3\% confidence level
from the likelihood distribution of the branching fraction. By
comparing the difference of the log-likelihood with and without the
$\dsssm$ signal in the fit and considering the change of the
number of degrees of freedom, the statistical significance of the
$\dsssm$ signal is estimated as $5.8\sigma$. The
mass of $\dsssm$ is measured to be ($2318.3\pm 1.2$)~MeV/$c^2$.

\begin{figure}
\incfig{0.23}{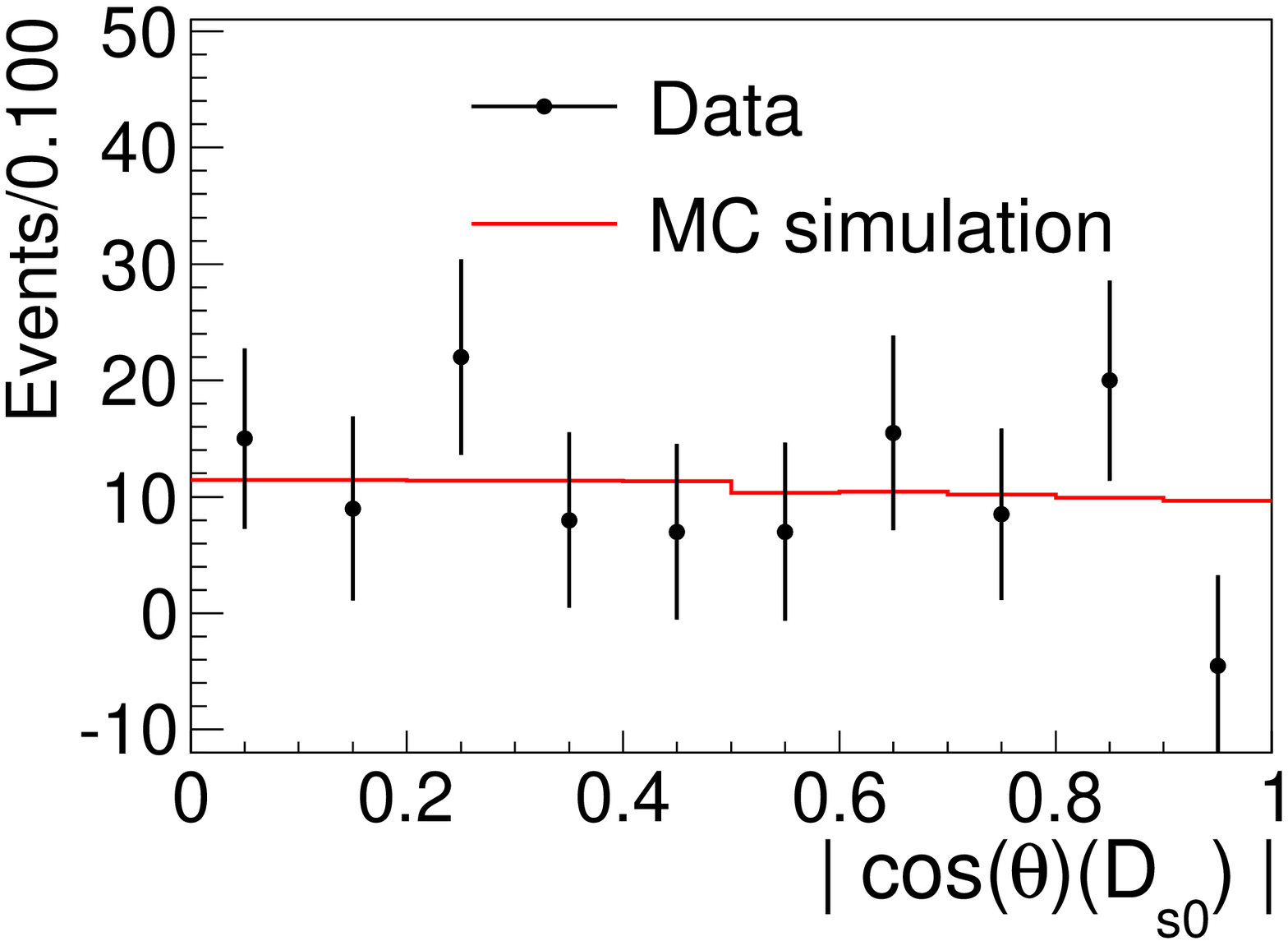}
\incfig{0.23}{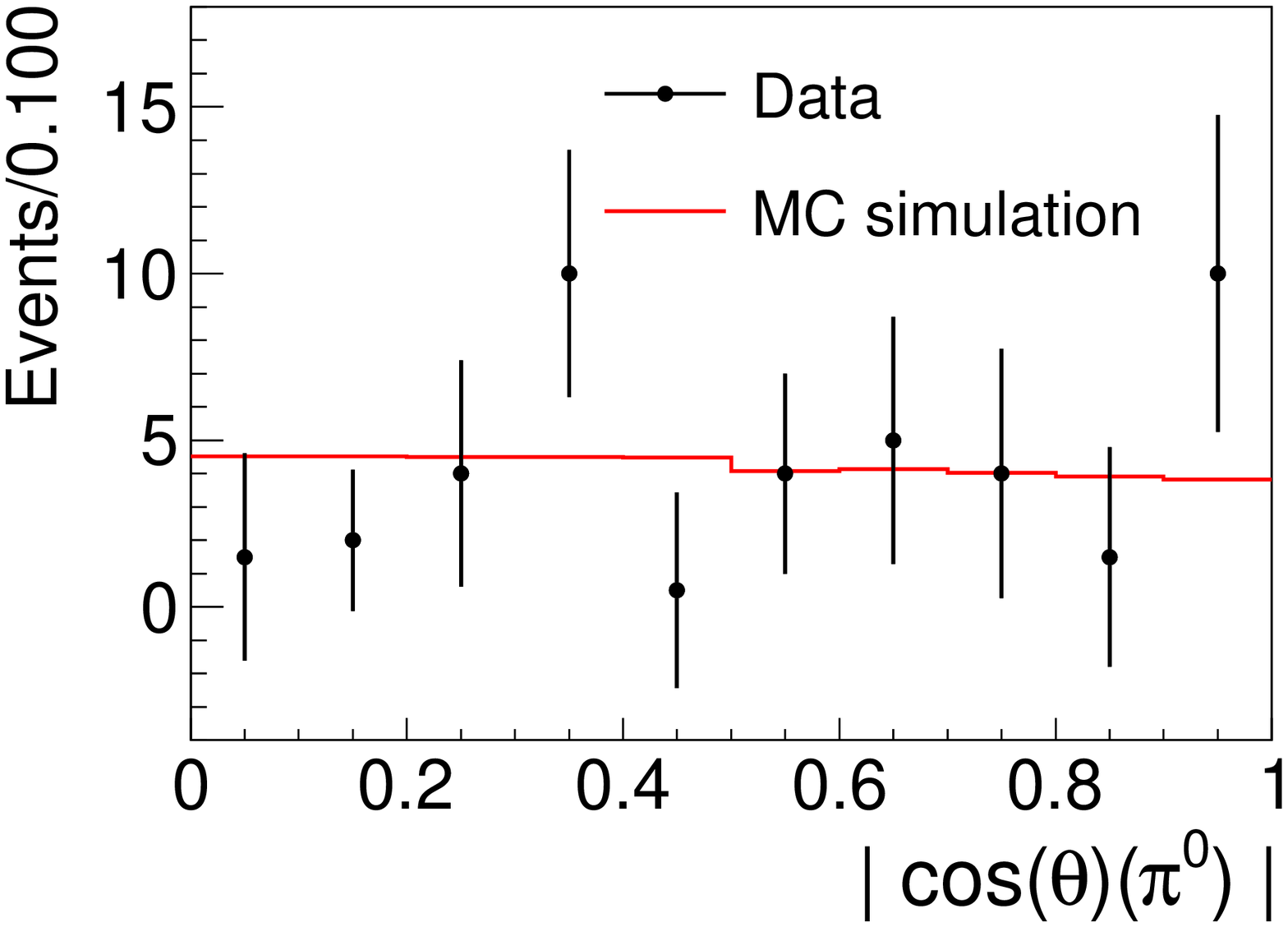}\\
\caption{(Color online) Angular distributions of $\dsssm$ in the $\EE$ c.m.\ system
(left) and of $\pi^{0}$ in the $\dsssm$ c.m.\ system (right). Black dots
and red lines represent the data after background subtraction and
MC simulation, respectively.} \label{fig:ang}
\end{figure}

The $J^{P}$ of $\dsss$ is $0^{+}$, so both the $\dssp\dsssm$ and
the $\pidsm$ systems are expected to be in a relative $S$-wave, and the angular
distributions are expected to be flat. We define the signal region of
$\dsssm$ as [2.31, 2.33]~GeV/$c^{2}$, and the sideband regions as
[2.28, 2.30] and [2.34, 2.36]~GeV/$c^{2}$ to estimate the
contribution of background. Figure~\ref{fig:ang} shows the angular
distributions of $\dsssm$ in the $\EE$ c.m.\ system and of $\pi^{0}$ in
the $\dsssm$ c.m.\ system.  Both distributions are flat as expected,
and can be modeled by the MC simulations.

For the branching fraction measurement, many sources of systematic
uncertainties cancel since the branching fraction is
determined by the relative signal yields in the two subsamples.
The main systematic uncertainties come from $\pi^{0}$ reconstruction,
the used signal and background shapes, $\pidsm$ selections, the possible width of
$\dsssm$, and potential peaking backgrounds.

The uncertainty on $\pi^{0}$ reconstruction is taken as 0.7\%
from a study of $\psi(3686)\to J/\psi \pi^{0} \pi^{0}$ and $\EE\to\omega\pi^{0}$
by considering the momentum dependency of $\pi^{0}$.
In the nominal fit, the signal shape is
parameterized by a Crystal Ball function with a tail due to the ISR
effect. Given that the energy dependent cross sections of $\EE\to
\dssp\dsssm$ are not measured with high precision, the systematic
uncertainty should be studied conservatively. We vary the signal
shape to a Gaussian with all parameters free, and the relative difference in
the branching fractions, 5.0\%, is taken as systematic
uncertainty. The background in the nominal fit is parameterized as
a linear function.  We change this shape to a second order
polynomial function and take the relative difference in branching fractions, 7.4\%, as
systematic uncertainty due to background shape.

For $\pidsm$ selection, we perform a kinematic fit, which could
cause a systematic bias in the efficiency between data and MC
simulation. To study this difference, we correct the helix
parameters of the charged tracks in MC simulation~\cite{pull1}, the
difference in $\chi^{2}$ distribution between data and MC
simulation becomes negligibly small according to other
studies~\cite{pull2}. We take half of the difference in the ratio of detection efficiencies $\epsilon_{\rm sig}$ and $\epsilon_{\rm tot}$ between MC
simulations with and without this correction as systematic
uncertainty (3.1\%).  The nominal result is based on
the corrected MC simulation.

The width of $\dsss$ is unknown and cannot be measured in
this analysis due to limited statistics. In the nominal fit,
we use the shape from MC simulation of $\dsssm$ with a zero width
 to describe the signal.
The upper limit on the width of $\dsssm$
is estimated as 3.8~MeV at 95\% C.L. from previous
experiments~\cite{pdg}. In an alternative fit, we change the width of $\dsssm$ to 3.8~MeV
and use the same Gaussian function to convolve the shape from MC simulation,
and take the difference in the branching fraction, 5.3\%, as
systematic uncertainty.

In Eq.~(\ref{equ:br}), the peaking background is considered, and
the result of the fit shows that its contribution is negligible.
For the signal mode,
$\dsssm\to \pidsm$, the tagged $\pi^{0}$ could also come from $\dsm$.
This kind of events is regarded as signal, and its contribution is included in
the definition of the efficiency, which is estimated from the MC simulation of $\EE\to
\dssp\dsssm\to \dssp\pidsm$ with $\dsm$ decaying to all possible modes.
All peaking backgrounds come from other decay modes of
$\dsssm$. To study the possible contribution conservatively, we
simulate the potential peaking backgrounds, $\dsssm\to
\gamma\dssm$, $\gamma\gamma \dsm$ and $\pipi \dsm$ exclusively.
The upper limits on the ratio $\Gamma(\gamma\dssm)/\Gamma(\pidsm)$,
$\Gamma(\gamma\gamma\dsm)/\Gamma(\pidsm)$, and
$\Gamma(\pipi\dsm)/\Gamma(\pidsm)$, are estimated as 0.059, 0.18, and
0.006~\cite{pdg}. The total systematic uncertainty in
$\mathcal{B}(\dsssm\to \pidsm)$ is conservatively estimated to be 8.5\%.

All the above systematic uncertainties are listed in
Table~\ref{tab:sys_err_br}. Assuming all of them are independent and adding them in quadrature,
we estimate a total systematic uncertainty of 13.8\% in the branching fraction.

\begin{table}[htbp]
\caption{\small Summary of relative systematic uncertainties in
$\mathcal{B}(\dsssm\to \pidsm)$.} \label{tab:sys_err_br}
\begin{tabular}{c c c}
\hline \hline
    Source               & Uncertainty (\%)\\
    \hline
    $\pi^{0}$ reconstruction     & 0.7  \\
    Signal shape         & 5.0  \\
    Background shape     & 7.4  \\
    $\pidsm$ selections   & 3.1  \\
    Width of $\dsssm$     & 5.3  \\
    Peaking backgrounds  & 8.5  \\
    \hline
    Total                & 13.8 \\
    \hline \hline
\end{tabular}
\end{table}

The systematic uncertainties in the mass measurement of $\dsssm$ come
from mass calibration, signal shape, background shape, and c.m.\
energy determination. For the mass calibration, we use the control sample $\EE\to
D_{s}^{*+} D_{s}^{*-}$ at 4.6~GeV and compare the mass of the
recoiling $\dssm$ with the world average value~\cite{pdg}. The same event
selections and fit procedure as for $\dssp\dsssm$ are used for
$\dssp\dssm$, and the shape of the missing $\dssm$ is parameterized as
a Crystal Ball function convolved with a Gaussian function. The difference
in the mass of $\dssm$ between data and the world average value~\cite{pdg},
which includes the contribution of the uncertainty on c.m.\ energy,
1.2~MeV/$c^{2}$, is taken as systematic uncertainty.
The uncertainties in signal and background shapes are studied with the same
method as for the systematic uncertainty study in branching
fraction measurement. The results show that these systematic
uncertainties are negligible.

In summary, we observe the $\dsssm$ signal in the process $\EE\to
\dssp\dsssm$ from a data sample at c.m.\ energy of 4.6~GeV. The statistical significance of $\dsssm$ signal is 5.8$\sigma$, and the mass is
determined to be ($2318.3\pm 1.2\pm 1.2$)~MeV/$c^{2}$. The absolute branching
fraction of $\dsssm\to \pidsm$ is measured for the first time to be $1.00^{+0.00}_{-0.14}\pm
0.14$, where the uncertainties are statistical
and systematic, respectively.
The result shows that the $\dsssm$ tends to have a significantly smaller branching fraction
to $\gamma\dssm$ than to $\pidsm$, and this differs from the expectation
of the conventional $\bar{c}{s}$ hypothesis of the $\dsssm$~\cite{br_model}
but agrees well
with the calculation in the molecule picture~\cite{br_molecule}.
In the future, with more data accumulated at BESIII
or a fine scan from PANDA~\cite{panda}, the width of
$\dsssm$ could be measured. Combined with the absolute branching fractions
of $\dsssm\to\pidsm$ and $\gamma\dssm$, we may shed light on the
nature of the $\dsssm$.

The BESIII collaboration thanks the staff of BEPCII and the IHEP computing center for their strong support. This work is supported in part by National Key Basic Research Program of China under Contract No. 2015CB856700; National Natural Science Foundation of China (NSFC) under Contracts Nos. 11235011, 11322544, 11335008, 11425524, 11635010; the Chinese Academy of Sciences (CAS) Large-Scale Scientific Facility Program; the CAS Center for Excellence in Particle Physics (CCEPP); the Collaborative Innovation Center for Particles and Interactions (CICPI); Joint Large-Scale Scientific Facility Funds of the NSFC and CAS under Contracts Nos. U1632106, U1232201, U1332201, U1532257, U1532258; CAS under Contracts Nos. KJCX2-YW-N29, KJCX2-YW-N45; 100 Talents Program of CAS; National 1000 Talents Program of China; INPAC and Shanghai Key Laboratory for Particle Physics and Cosmology; German Research Foundation DFG under Contracts Nos. Collaborative Research Center CRC 1044, FOR 2359; Istituto Nazionale di Fisica Nucleare, Italy; Koninklijke Nederlandse Akademie van Wetenschappen (KNAW) under Contract No. 530-4CDP03; Ministry of Development of Turkey under Contract No. DPT2006K-120470; National Natural Science Foundation of China (NSFC) under Contract No. 11575133; National Science and Technology fund; NSFC under Contract No. 11275266; The Swedish Resarch Council; U. S. Department of Energy under Contracts Nos. DE-FG02-05ER41374, DE-SC-0010504, DE-SC0012069; University of Groningen (RuG) and the Helmholtzzentrum fuer Schwerionenforschung GmbH (GSI), Darmstadt; WCU Program of National Research Foundation of Korea under Contract No. R32-2008-000-10155-0; New Century Excellent Talents in University (NCET) under Contract No. NCET-13-0342; Shandong Natural Science Funds for Distinguished Young Scholar under Contract No. JQ201402.

\end{document}